\documentclass[a4paper,11pt]{article}
\usepackage{pos}
\usepackage{subcaption}
\usepackage[switch]{lineno}
\newcommand{\gammaray}{$\gamma$-ray~}
\newcommand{\gammarays}{$\gamma$-rays~}

\title{Standardized Formats for Gamma-Ray Analysis Applied to HAWC Observatory Data}
 \ShortTitle{Standardized Formats for HAWC}

\manuallySeparateAuthors
\author*[a]{Laura Olivera-Nieto}
\author[b]{, Vikas Joshi}
\author[a]{, Harm Schoorlemmer}
\author[a]{ and Axel Donath}

\affiliation[a]{Max Planck Institut f\"{u}r Kernphysik,\\
  Saupfercheckweg 1, 69117 Heidelberg, Germany}

\affiliation[b]{Erlangen Centre for Astroparticle Physics,\\
Erwin-Rommel-Strasse 1, 91058 Erlangen, Germany}

\forColl{HAWC} 

\emailAdd{laura.olivera-nieto@mpi-hd.mpg.de}

\abstract{A wide range of data formats and proprietary software have traditionally been used in \gammaray astronomy, usually developed for a single specific mission or experiment. However, in recent years there has been an increasing effort towards making astronomical data open and easily accessible. Within the \gammaray community this has translated to the creation of a common data format across different \gammaray observatories: the "gamma-astro-data-format" (GADF). Based on a similar premise, open-source analysis packages, such as Gammapy, are being developed and aim to provide a single, robust tool which suits the needs of many experiments at once. In this contribution we show that data from the High-Altitude Water Cherenkov (HAWC) observatory can be made compatible with the GADF and present the first GADF-based production of event lists and instrument response functions for a ground-based wide-field instrument. We use these data products to reproduce with excellent agreement the published HAWC Crab spectrum using Gammapy. Having a common data format and analysis tools facilitates joint analysis between different experiments and effective data sharing. This will be especially important for next-generation instruments, such as the proposed Southern Wide-field Gamma-ray Observatory (SWGO) and the planned Cherenkov Telescope Array (CTA).}

\FullConference{37$^{\rm{th}}$ International Cosmic Ray Conference (ICRC 2021)\\
		July 12th -- 23rd, 2021\\
		Online -- Berlin, Germany}

\begin{document}
\maketitle

\section{Introduction}
\label{sec:introduction}
 Historically, a variety of instrument-specific and largely proprietary tools and data formats have been used in \gammaray astronomy, which hinders effective data-sharing and reproducibility. However, in recent years there has been a shift towards making data more accessible and easier to share in the context of joint analysis. 
 A big driver of this trend has been the upcoming Cherenkov Telescope Array (CTA)~\cite{CTA}, the data of which will become public after a short proprietary period. Motivated by this development, there has been an advent of openly developed analysis tools, such as \textit{Gammapy}~\cite{gammapy-icrc17}
, which are able to replace the existing instrument-specific packages by offering a single, common tool. \textit{Gammapy} is a community developed Python package for $\gamma$-ray astronomy selected to be part of the CTA science tools. It has been successfully used and validated for analysis of Imaging Atmospheric Cherenkov Telescope (IACT) data~\cite{lars} and joint analysis with data from the Fermi Large Area Telescope~\cite{cosimo}. 

In parallel to these efforts, a common data format across different observatories, the \textit{gamma-astro-data-format} (GDAF)\footnote{\url{https://gamma-astro-data-formats.readthedocs.io/en/latest/}}~\cite{gadf-icrc17} has been developed. The scope of this standard is to cover all high-level data products from telescopes, starting at the level of event lists and instrument response functions (IRFs). This format relies on file storage by the Flexible Image Transport System (FITS) format \cite{fits}, which is widely used by the whole astronomical community. It builds on existing standards such as OGIP\footnote{\url{https://heasarc.gsfc.nasa.gov/docs/heasarc/ofwg/ofwg\_intro.html}} and expands them to tailor the specific needs of the \gammaray community. The focus of these efforts has largely been IACT data, ignoring another type of \gammaray experiment: wide-field ground-based observatories.

The High-Altitude Water Cherenkov (HAWC) is a \gammaray detector that consists of 300 water Cherenkov detectors, each outfitted with four photomultiplier tubes (PMTs). Its wide field of view covers two-thirds of the sky uninterruptedly, allowing for constant monitoring and deep observations of the \gammaray sky. It has been in operation since November 2014.

It is worth noting that the HAWC Accelerated Likelihood (HAL)~\cite{hawc-hal} framework and the Multi-Mission Maximum Likelihood framework (3ML)~\cite{3ml}, the packages primarily used by the HAWC observatory, are also open-source, but spouse a different philosophy to that of the packages described above. Instead of replacing the existing frameworks from different observatories by a single, common tool, packages like 3ML provide a common framework in which the instrument-specific tools (such as HAL) interface. This approach has the advantage to include not only \gammaray instruments but also multiwavelength and multimessenger observations. However, it has the disadvantage to rely on instrument-specific analysis software, which, in the case of \gammaray observations, performs very similar tasks between different instruments. 

\section{HAWC data and IRFs in the GADF}
\label{sec:dl3}
Events recorded by the HAWC observatory are binned by size, that is, on the fraction of PMTs from the array that were triggered by the event. A total of 9 event size bins are defined, as detailed in Table~\ref{table:nhitbins}. The event size only weakly correlates with energy~\cite{hawc-crab-nhit}. In order to estimate the energy on an event-by-event basis, more advanced algorithms have been developed. The ground parameter (GP) algorithm is based on the charge density deposited at ground by the shower. The neural network (NN) algorithm estimates energies with an artificial neural network that takes as input several quantities computed during the event reconstruction. A detailed overview of both algorithms can be found in~\cite{hawc-crab}. Energy bins are typically defined beforehand in HAWC analysis and referred to with letter names, as described in Table~\ref{table:energybins}. The combination of event sizes and energy bins leads to a 2-dimensional bin scheme, with 108 analysis bins resulting from the combination of each event size bin 1 to 9 with the 12 energy bins. However, only a subset of these bins are populated with enough event statistics, for example, low energy events are very unlikely to have large event sizes.

\begin{table}
	\small
	\parbox{.48\linewidth}{
		\centering
		\scalebox{0.85}{
	\begin{tabular}{|c|c|c|}
		\hline
		Bin number & Low edge & High edge\\
		\hline
		1 & 0.067 & 0.105 \\
		2 & 0.105 & 0.162 \\
		3 & 0.162 & 0.247 \\
		4 & 0.247 & 0.356 \\
		5 & 0.356 & 0.485 \\
		6 & 0.485 & 0.618 \\
		7 & 0.618 & 0.740 \\
		8 & 0.740 & 0.840 \\
		9 & 0.840 & 1.00  \\
		\hline
	\end{tabular}}
	\caption{Event size ("\textit{nhit}") bins. Bins are defined from the fraction of PMTs triggered by each event.}
		\label{table:nhitbins}
	}
	\hfill
	\parbox{.48\linewidth}{
		\centering
				\scalebox{0.85}{
	\begin{tabular}{|c|c|c|}
		\hline
		Bin & Low edge (TeV) & High edge (TeV)\\
		\hline
		a & 0.316 &   0.562 \\
		b & 0.562 &   1.00  \\
		c & 1.00  &   1.78  \\
		d &   1.78  &   3.16  \\
		e &   3.16  &   5.62  \\
		f &   5.62  &  10.0   \\
		g &  10.0   &  17.8   \\
		h &  17.8   &  31.6   \\
		i &  31.6   &  56.2   \\
		j &  56.2   & 100     \\
		k & 100     & 177     \\
		l & 177     & 316     \\
		\hline
	\end{tabular}}
	\caption{The energy bins. Note that the first two bins are not used in the analysis as the estimate is highly biased~\cite{hawc-crab}.}
	\label{table:energybins}
	}
\end{table}

\subsection{Events and good time intervals}
For each of these analysis bins, gamma-hadron separation cuts are optimized~\cite{hawc-crab} and applied to the reconstructed data. Additional direction corrections are applied and the event coordinates are converted to the J2000 epoch. The resulting event lists are stored in FITS~\cite{fits} files with headers and columns compliant with the GADF. Additionally, other columns are stored that contain information pertinent to the characteristics of wide-field arrays, such as core location in the array, direction in local coordinates and event size.

Due to the large sky coverage and high duty cycle, it is common among wide-field observatories to make sky-maps as their primary data product directly from reconstructed events. However, producing event lists as an intermediate step has the advantage of adding more flexibility. It allows to easily select a subset of the whole dataset for a given analysis, which simplifies the study of time-dependent signals. It also facilitates extensive systematic checks.

\begin{figure}
	\centering
	\includegraphics[width=.55\textwidth]{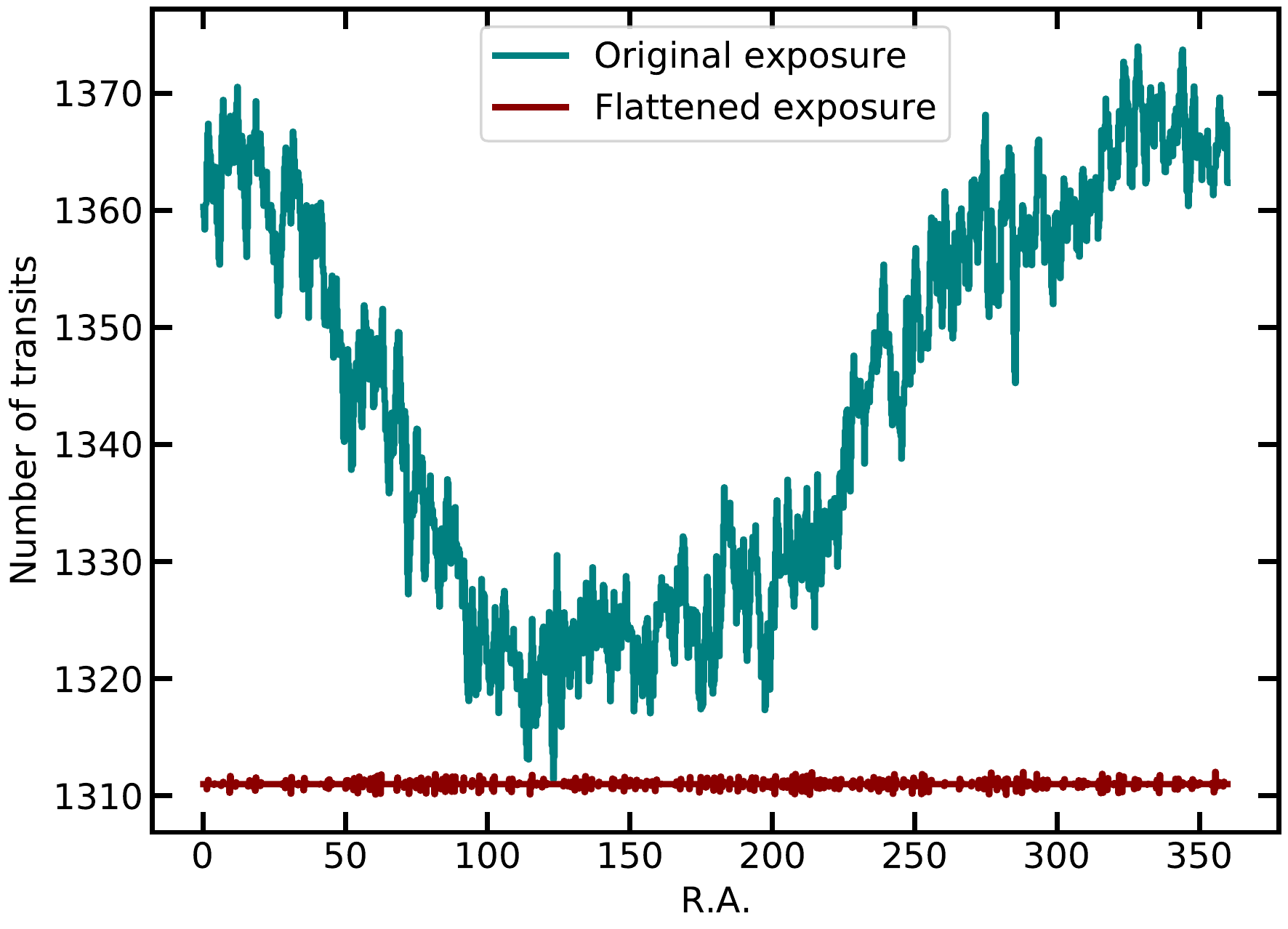}
	\caption{Number of transits during which data was recorded as a function of R.A.}
	\label{fig:exposure}
\end{figure}

Good Time Intervals (GTIs) are defined as the time intervals during which the detector is on and taking data continuously. These intervals are then used to build the exposure information. Detector downtime is caused by a variety of factors, ranging from hardware issues to meteorological conditions. These interruptions are not uniformly distributed over time, with certain meteorological events being more likely during particular times of the year. This leads to fluctuations in the exposure as a function of hour angle, or, equivalently, right ascension (R.A.). For HAWC data collected between June 2015 and June 2019, we compute the number of transits - i.e., sidereal days - for which the detector was taking data as a function of the R.A. of zenith, shown by the green line in Figure~\ref{fig:exposure}.
It can be useful, for example for background modeling, to remove this R.A. dependency on the exposure, to "flatten" it. To do this, one can remove GTIs from the time selection until the resulting exposure is no longer a function of R.A. An example of this process can be seen in Figure~\ref{fig:exposure} This exposure "flattening" leads to a loss of data of around 1-2\%, maintaining the overall detector efficiency above 90\%.

\subsection{Intrument response functions}
The IRFs describe the combined detection abilities and precision of an instrument data-taking and reconstruction procedure. They are computed by simulating a point source emitting \gammarays following a given energy spectrum, usually $\propto$E$^{-2}$. These events are processed with the detector simulation and reconstruction procedure, see~\cite{hawc-crab-nhit} for more details. The reconstructed events are binned as described in Section~\ref{sec:dl3} and gamma-hadron separation is applied. This process yields information on the number of events successfully identified as \gammarays as well as the accuracy and precision of the energy and direction assigned to them. In the GADF, this information is split into three components respectively, the effective area, the energy dispersion matrix, and the point-spread function (PSF). Note that this framework
neglects  the  correlation  between the different IRF components. This  is  currently  mostly sufficient for IACT analysis and also present in the standard HAWC framework. However, this will be re-addressed for CTA and thus, likely for the GADF as well in future. 
\subsubsection{Effective area }
\begin{figure} [h!]
	\centering
	\begin{subfigure}[b]{0.48\textwidth}
		\centering
		\includegraphics[width=\textwidth]{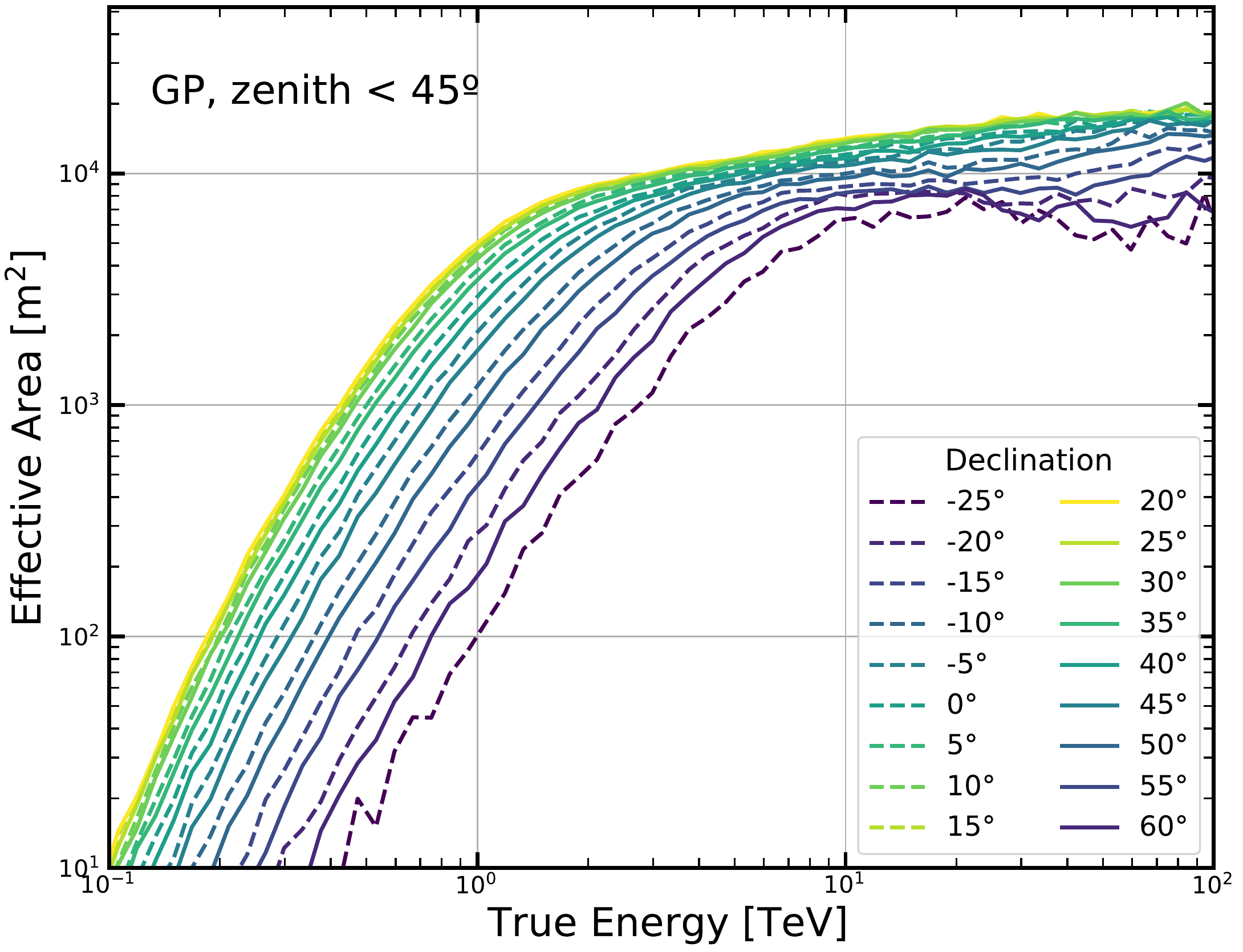}
	\end{subfigure}
	\hfill
	\begin{subfigure}[b]{0.48\textwidth}
		\centering
		\includegraphics[width=\textwidth]{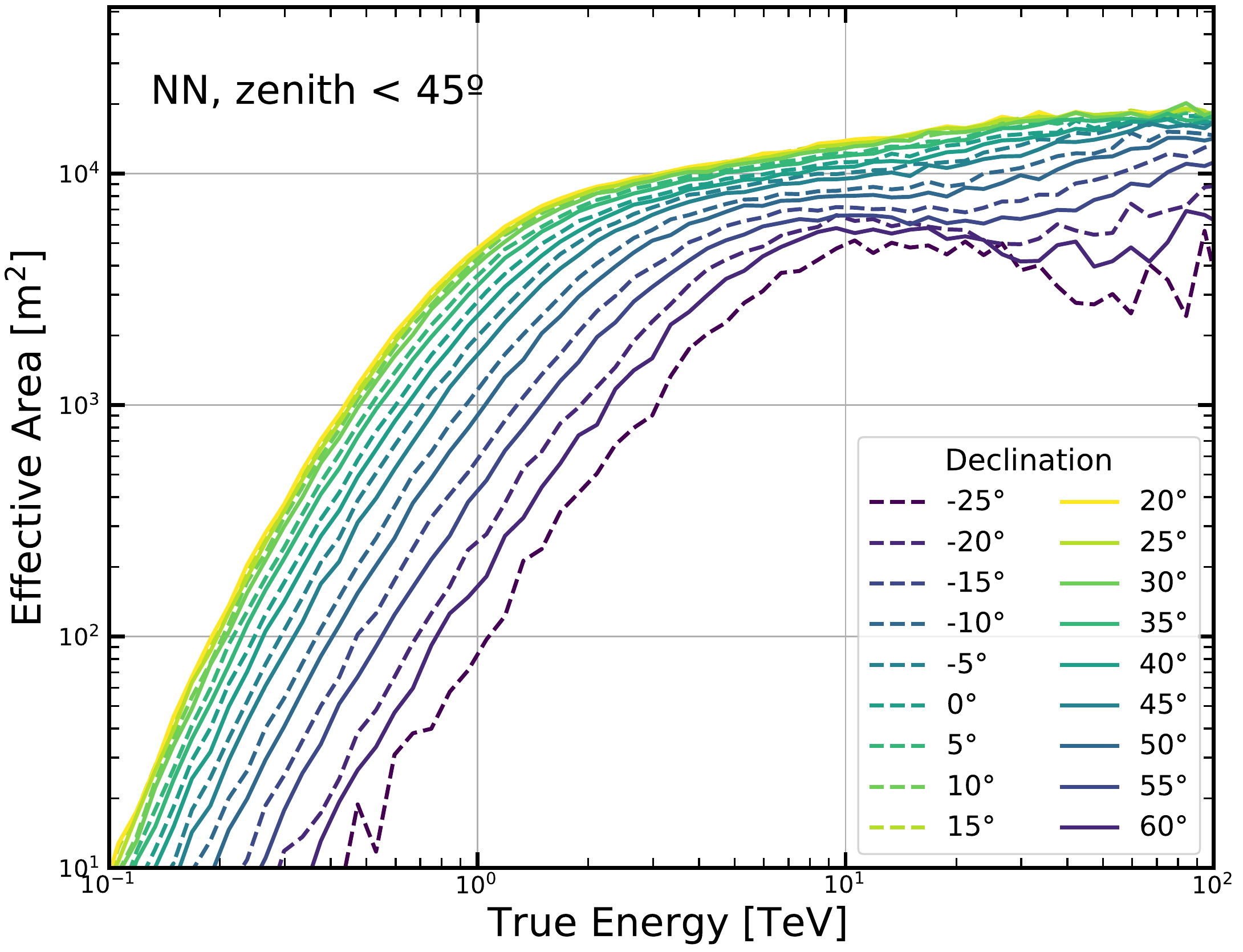}
	\end{subfigure}
	
	\caption{Effective area after background rejection for bins defined using each of the energy estimators described in Section~\ref{sec:dl3}.}
	\label{fig:aeff}
\end{figure}

The effective area of a detector is the combination of its detection efficiency with the observable
area. Typically in the computation of HAWC IRFs the simulated spectrum is convolved with a source transit over the observatory, leading to a comparison of events detected to events thrown per square meter and transit. 
Analysis bins described in Section~\ref{sec:dl3} include a cut on zenith angle at 45\textdegree~\cite{hawc-crab}. 
We can compute the number of hours that a source at each declination spends at zenith angles lower than 45\textdegree, that is, the duration of a transit as a function of declination, to recover the effective area in units of m$^2$. Figure~\ref{fig:aeff} shows these curves for all the available declinations resulting from analysis bins defined using each of the energy estimation schemes described in Section~\ref{sec:dl3}. The effective is greatest for declinations close to 19\textdegree, the terrestrial longitude of the HAWC location. Sources at this declination pass through the local zenith as they transit the sky over the observatory.  The effective area per transit, that is, the effective exposure for one transit, is also a useful quantity, especially when considering the long-term study of sources. Given its declination dependence, we fill a \textit{Gammapy} \texttt{Map}\footnote{\url{https://docs.gammapy.org/0.18.2/maps/index.html}} with the one-transit effective exposure. Then, for a given data range selection, the number of transits like the one shown in Figure~\ref{fig:exposure} is combined with this map to produce the effective exposure map used for the analysis.

\subsection{Point-Spread Function}
\begin{figure}[h!]
	\centering

	\includegraphics[width=0.95\textwidth]{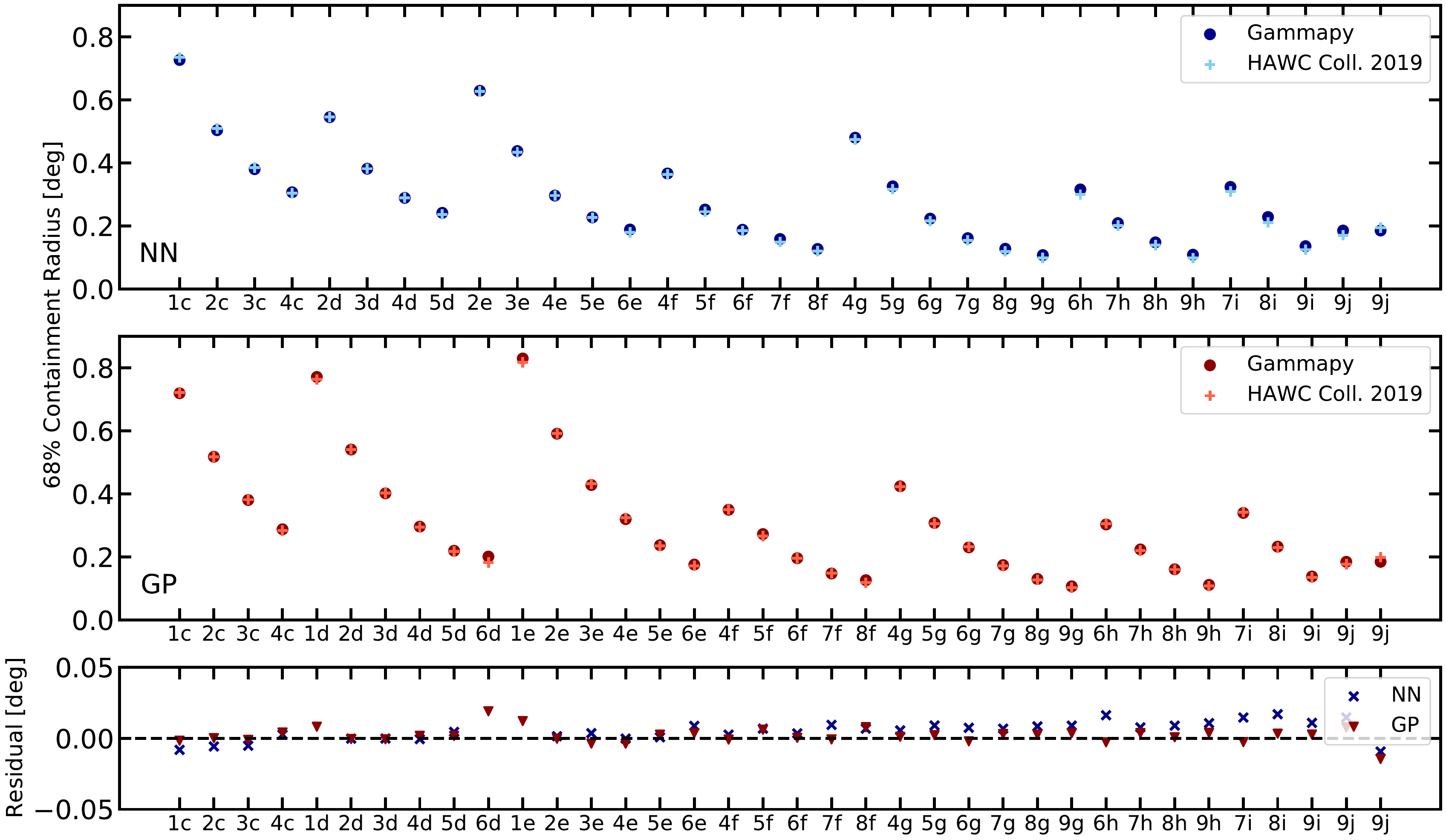}
	\caption{Comparison of the PSF containment at the Crab location for analysis bins using the energy estimators described in Section~\ref{sec:dl3}.}
	\label{fig:PSF}
\end{figure}

The point-spread function is a measure of the precision achieved in the event direction reconstruction. It is computed as the spatial probability distribution of events produced by a point source. It is assumed to be radially symmetric, and so can be stored as a function of offset from the source location. \textit{Gammapy} provides several options to store and use this information in the \texttt{irf}\footnote{\url{https://docs.gammapy.org/0.18.2/irf/psf.html\#irf-psf}} class. Like the other IRFs, the HAWC PSF depends on declination. For this reason, we fill a \texttt{PSFMap} with the PSF radial profile information at each sky position. From this, we can compute the 68\% containment radius, which we compare to the published values in~\cite{hawc-crab} to validate the procedure. This comparison can be seen in Figure~\ref{fig:PSF}, showing good agreement.
\subsection{Energy dispersion matrix}
\begin{figure} [h!]
	\centering
	\begin{subfigure}[b]{0.48\textwidth}
		\centering
		\includegraphics[width=\textwidth]{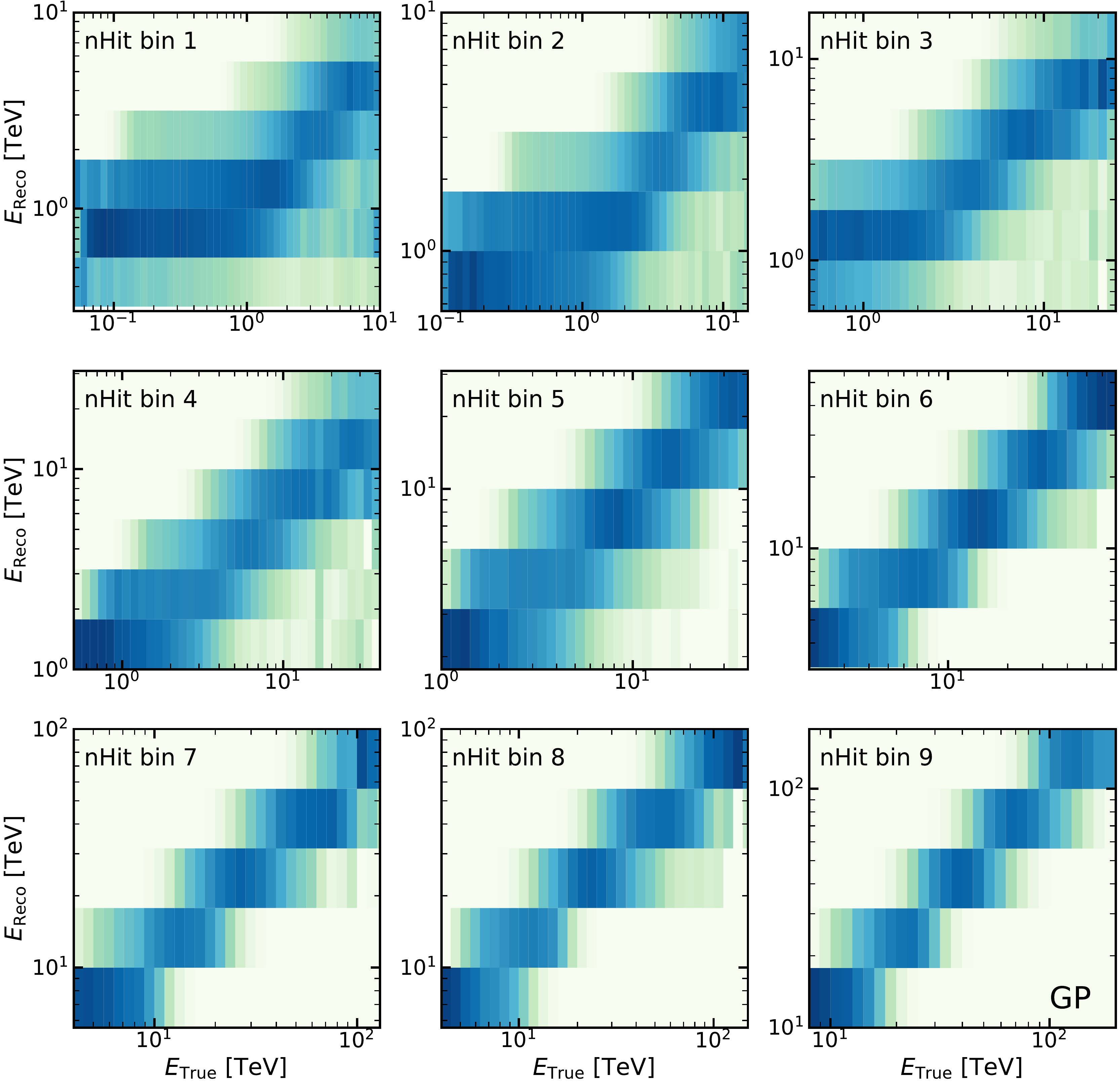}
	\end{subfigure}
	\hfill
	\begin{subfigure}[b]{0.48\textwidth}
		\centering
		\includegraphics[width=\textwidth]{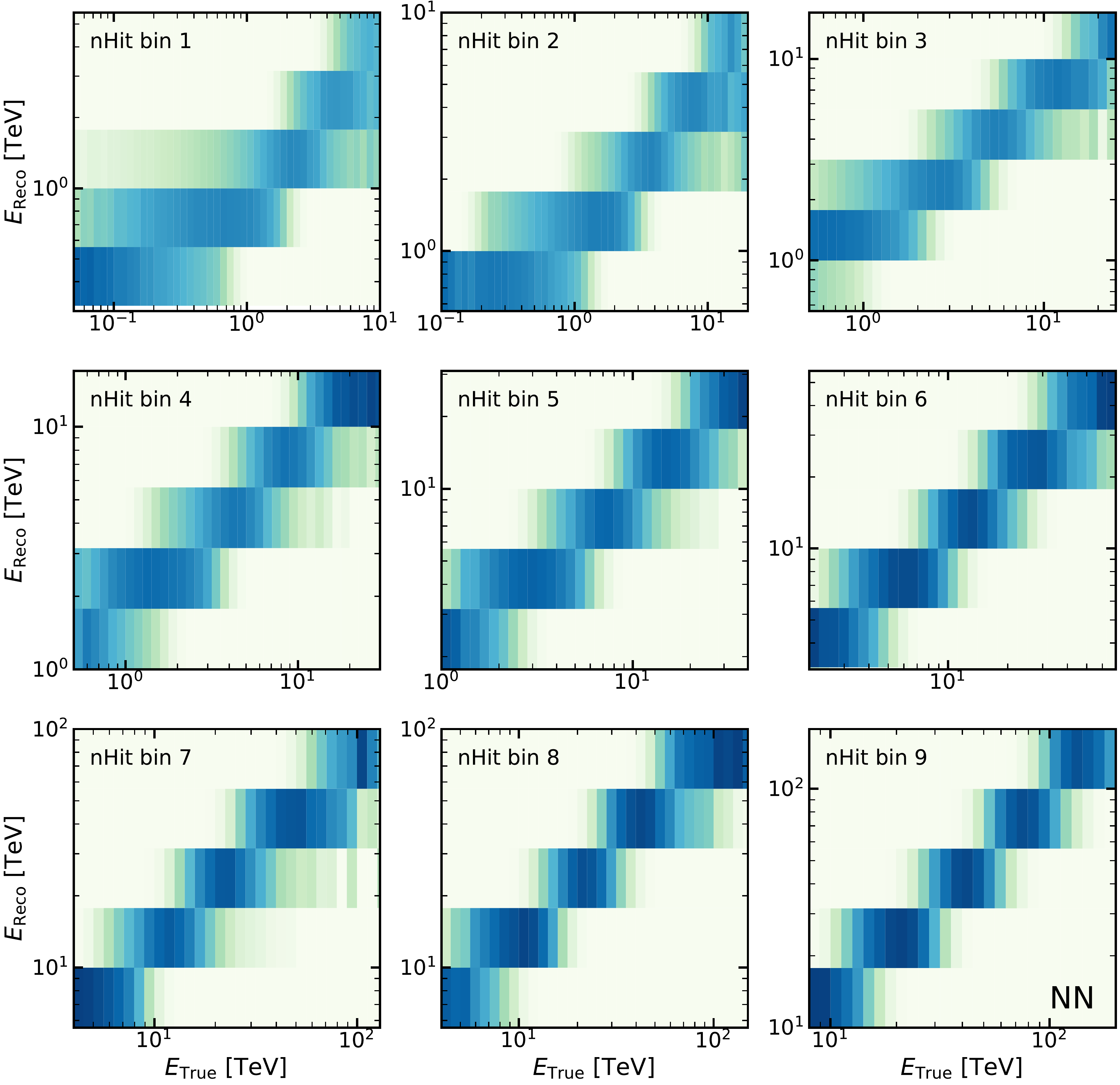}
	\end{subfigure}

	\caption{Energy dispersion matrix at the Crab location for each of the energy estimators described in Section~\ref{sec:dl3}}
		\label{fig:edisp}
\end{figure}

The energy dispersion is a measure of the accuracy and precision achieved in the event energy estimation. It is computed as the probability of an event with a given simulated energy (E$_{\mathrm{true}}$) to be reconstructed with a different energy (E$_{\mathrm{reco}}$). We use the \texttt{EDispMap}\footnote{\url{https://docs.gammapy.org/0.18.2/api/gammapy.irf.EDispMap.html\#gammapy.irf.EDispMap}} class in \textit{Gammapy}, which is a dedicated 4-dimensional sky-map. At each sky position, it contains the probability matrix that quantifies the energy dispersion. An example of such matrices at the Crab location is shown in Figure~\ref{fig:edisp} for each of the event size bins defined in Table~\ref{table:nhitbins} and E$_{\mathrm{reco}}$ energy bins defined in Table~\ref{table:energybins}. As can be seen, the energy resolution improves with increasing event size bin.

\section{Validation analysis with Gammapy: the Crab Nebula}
\label{sec:gammapy}

\begin{figure} [h!]
	\centering
	\begin{subfigure}[b]{0.48\textwidth}
		\centering
		\includegraphics[width=\textwidth]{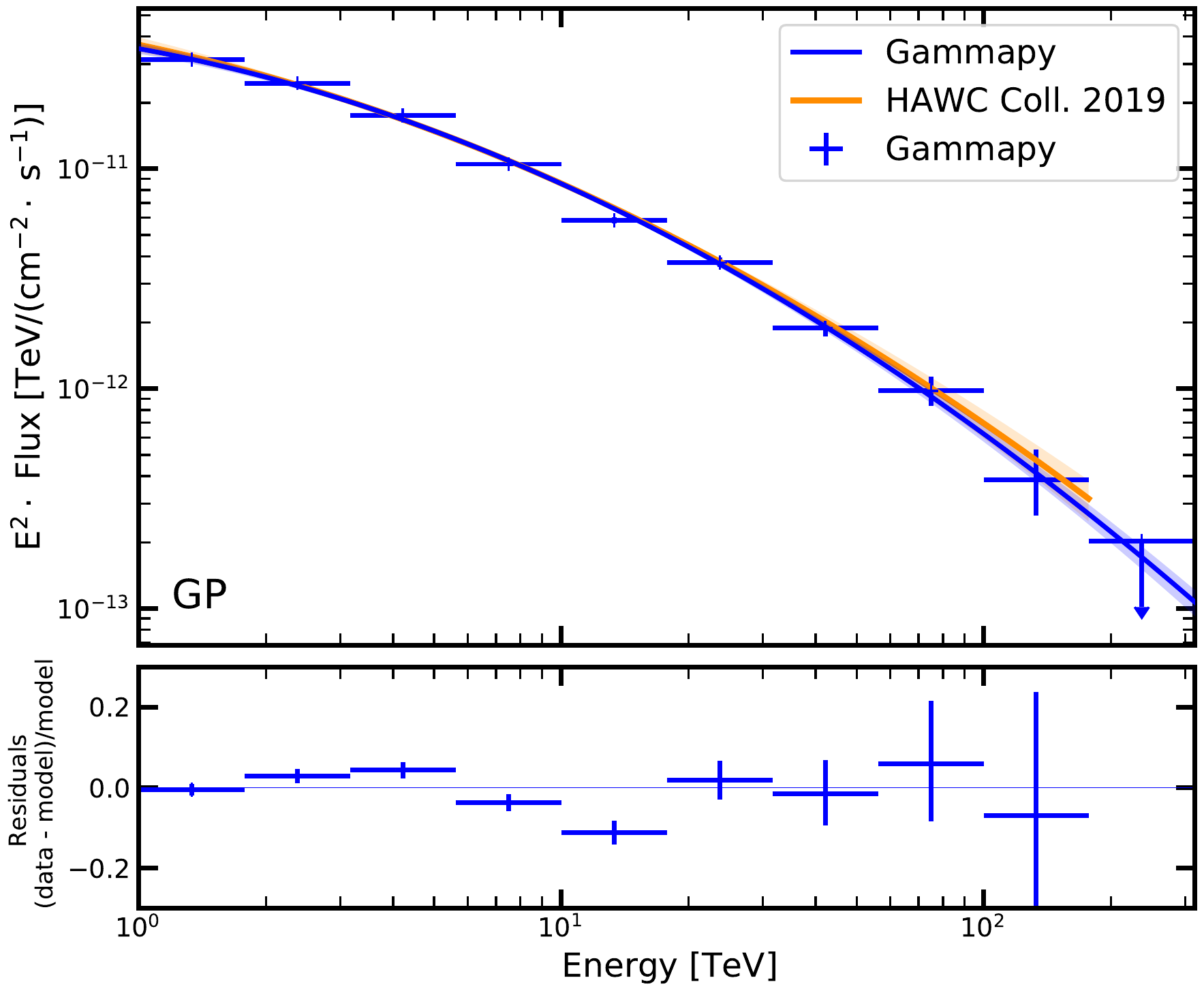}
	\end{subfigure}
	\hfill
	\begin{subfigure}[b]{0.48\textwidth}
		\centering
		\includegraphics[width=\textwidth]{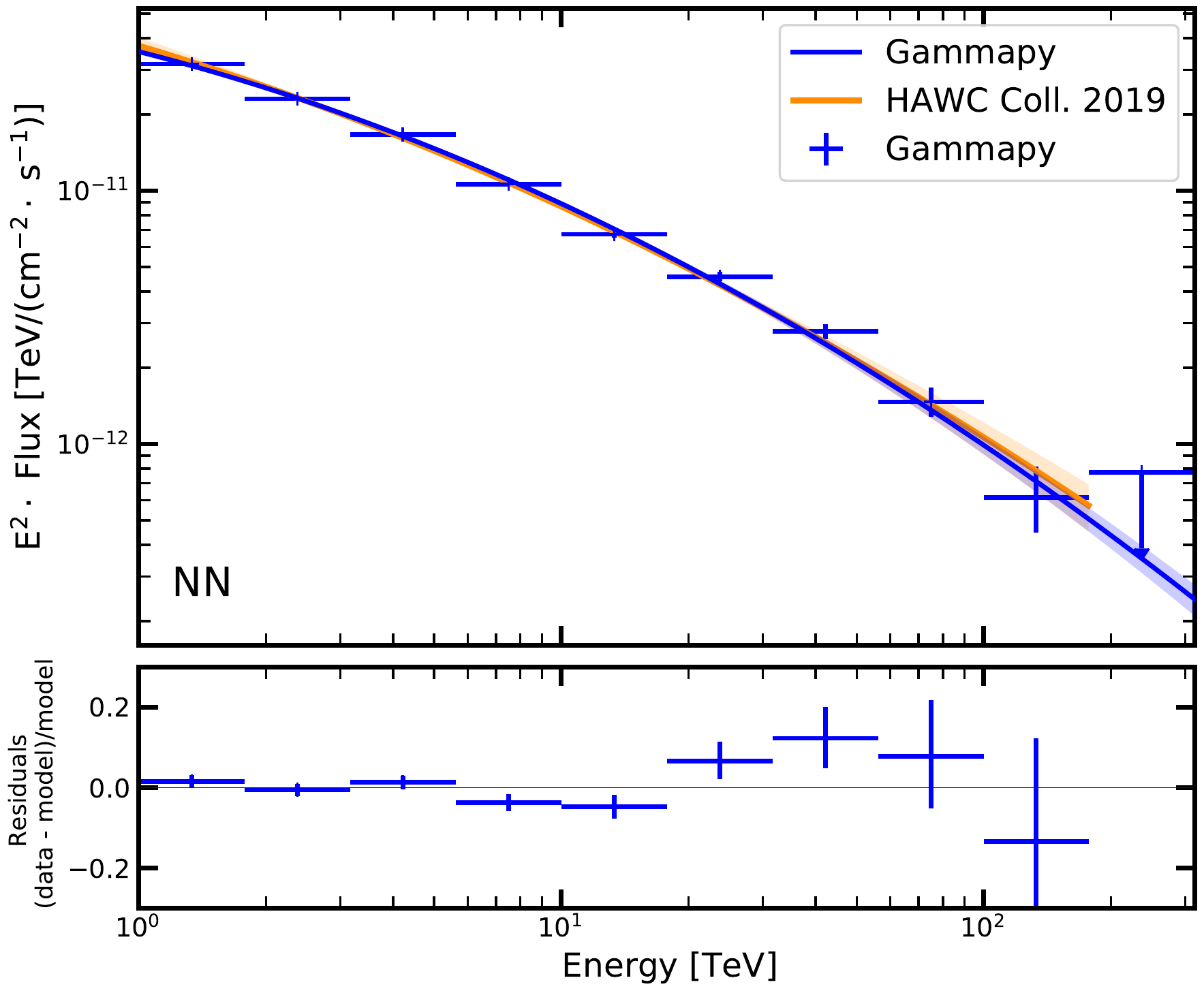}
	\end{subfigure}
	
	\caption{ Best-fit Crab spectrum obtained with \textit{Gammapy} compared with~\cite{hawc-crab} for both energy estimators described in the text. The residual shows the comparison of the flux points with the model in  ~\cite{hawc-crab}.}
	\label{fig:crab}
\end{figure}

The Crab Nebula is one of the brightest sources in the \gammaray sky. For this reason, it is typically used for calibration and as a reference analysis. The HAWC Collaboration published a measurement of the Crab spectrum extending up to 100 TeV using the energy estimators described in~\cite{hawc-crab}. Using the same data range and background estimation method as in that work, and the IRFs as described above, we can reproduce that result using \textit{Gammapy}. We fit a combined 3-dimensional (spatial and spectral) model. For the spatial model we use a point-source assumption, and a log-parabola for the spectral shape. The result of this fit is shown in Figure~\ref{fig:crab}, for both energy estimators. For both figures, the top panel compares the best-fit spectrum obtained with \textit{Gammapy} with the published one, showing excellent agreement. In the bottom panel, the flux points computed with \textit{Gammapy} are compared to the model in~\cite{hawc-crab}.
\section{Conclusion}
\label{sec:conclusion}

The data from ground-based, wide-field observatories, and in this case, from the HAWC observatory, can also be made compatible with the GDAF and thus can be analyzed using the related open-source tools with minor adjustments. We find excellent agreement with the results published in~\cite{hawc-crab} using an analysis tool that is built with a different philosophy and structure. This is a powerful check of both the scientific result and both tools involved, as well as the production of HAWC data and IRFs in the GDAF-compliant format.

Having a common data format and analysis tools facilitates joint analysis between different experiments and effective data sharing. This synergy between experiments is particularly relevant given the complimentary nature of pointing and wide-field instruments. This will be specially important for future observatories like SWGO~\cite{swgo}.

The lifetime of observatories is finite, and one of the concerns at the end of the operation is to ensure that the archival data is available and easy to use for future studies and reproducibility of results. Having data in a format that is common to other observatories and which can be analyzed with a general use tool is a great advantage in this regard. 

\textit{Gammapy} has recently been selected as the official CTA Science tool. This ensures that it will be maintained and used by the overall \gammaray community, a much larger developer and user base than any of the other collaboration-specific tools individually.

\acknowledgments
\footnotesize
We acknowledge the support from: the US National Science Foundation (NSF); the US Department of Energy Office of High-Energy Physics; the Laboratory Directed Research and Development (LDRD) program of Los Alamos National Laboratory; Consejo Nacional de Ciencia y Tecnolog\'ia (CONACyT), M\'exico, grants 271051, 232656, 260378, 179588, 254964, 258865, 243290, 132197, A1-S-46288, A1-S-22784, c\'atedras 873, 1563, 341, 323, Red HAWC, M\'exico; DGAPA-UNAM grants IG101320, IN111716-3, IN111419, IA102019, IN110621, IN110521; VIEP-BUAP; PIFI 2012, 2013, PROFOCIE 2014, 2015; the University of Wisconsin Alumni Research Foundation; the Institute of Geophysics, Planetary Physics, and Signatures at Los Alamos National Laboratory; Polish Science Centre grant, DEC-2017/27/B/ST9/02272; Coordinaci\'on de la Investigaci\'on Cient\'ifica de la Universidad Michoacana; Royal Society - Newton Advanced Fellowship 180385; Generalitat Valenciana, grant CIDEGENT/2018/034; Chulalongkorn University’s CUniverse (CUAASC) grant; Coordinaci\'on General Acad\'emica e Innovaci\'on (CGAI-UdeG), PRODEP-SEP UDG-CA-499; Institute of Cosmic Ray Research (ICRR), University of Tokyo, H.F. acknowledges support by NASA under award number 80GSFC21M0002. We also acknowledge the significant contributions over many years of Stefan Westerhoff, Gaurang Yodh and Arnulfo Zepeda Dominguez, all deceased members of the HAWC collaboration. Thanks to Scott Delay, Luciano D\'iaz and Eduardo Murrieta for technical support.

\bibliographystyle{JHEP}
\bibliography{references} 

\clearpage
\section*{Full Authors List: HAWC Collaboration}

\scriptsize
\noindent
A.U. Abeysekara$^{48}$,
A. Albert$^{21}$,
R. Alfaro$^{14}$,
C. Alvarez$^{41}$,
J.D. \'Alvarez$^{40}$,
J.R. Angeles Camacho$^{14}$,
J.C. Arteaga-Vel\'azquez$^{40}$,
K. P. Arunbabu$^{17}$,
D. Avila Rojas$^{14}$,
H.A. Ayala Solares$^{28}$
% ...

\noindent
% ...
$^{14}$Instituto de F\'isica, Universidad Nacional Aut\'onoma de M\'exico, Ciudad de M\'exico, M\'exico,
% ...
$^{17}$Instituto de Geof\'isica, Universidad Nacional Aut\'onoma de M\'exico, Ciudad de M\'exico, M\'exico,
% ...
$^{21}$Physics Division, Los Alamos National Laboratory, Los Alamos, NM, USA,
% ...
$^{28}$Department of Physics, Pennsylvania State University, University Park, PA, USA,
% ...
$^{40}$Instituto de F\'isica y Matem\'aticas, Universidad Michoacana de San Nicol\'as de Hidalgo, Morelia, Michoac\'an, M\'exico,
$^{41}$FCFM-MCTP, Universidad Aut\'onoma de Chiapas, Tuxtla Guti\'errez, Chiapas, M\'exico,
% ...
$^{48}$Department of Physics and Astronomy, University of Utah, Salt Lake City, UT, USA,
% ...

\end{document}